\documentclass[11pt]{article}
\usepackage{amsmath,amsthm}
\usepackage{amssymb, latexsym}
\usepackage[mathscr]{euscript}
\usepackage{color}
\usepackage{array}
\usepackage{flafter}
\usepackage{layout}
\usepackage{graphicx}
\usepackage[applemac]{inputenc}
\usepackage{hyperref}
\usepackage{natbib}

\begin{document}

\def\spacingset#1{\renewcommand{\baselinestretch}%
{#1}\small\normalsize} \spacingset{1}

\title{\bf $F$ tests for the strip-split plot design}

\author{Daniel Andr\'es D\'{i}az-Pach\'on\thanks{Biostatistics Division -- University of Miami, DDiaz3@med.miami.edu} \and Francisco J. P. Zimmermann\thanks{EMBRAPA -- Rice and Bean, zimmermann.francisco@gmail.com} \and Luis Alberto L\'opez\thanks{Statistics Department -- Universidad Nacional de Colombia, lalopezp@unal.edu.co}}
\maketitle

\begin{abstract}
\noindent In this article we present the structure of the $F$ tests, the variance components and the approximate degrees of freedom for each of the eight possible mixed models of the strip-split plot design. We present an example to illustrate the model and compare it to more traditional settings like a three-way factorial design and a split-split plot model.
\end{abstract}


\section{Introduction and Method}
There are many opportunities in which a researcher needs to know the behavior of a factor in relation to one and/or two additional factors. When this happens it is usually considered a split plot design adding subplots into the former plots. It is due, in part, to the great development reached by this type of experiment.

Instead, we study here the strip-split-plot design; i.e., an extension of strip-block designs such that each plot on the intersection is subdivided into subplots to insert a third factor. This new factor will be more precise on its measurement due to its high number of observations and interactions; which is the more important feature of the design.

We do not claim originality on the invention of this model. On the contrary, \cite{Gomez} described it, as well as \cite{Zimm} did. They also described the $F$ tests when the effects are fixed. Nonetheless, after an intensive search, we could not find on the literature those same $F$ tests for the strip-split plot design with mixed effects. Our aim here is to fill that gap. On a sequel, we will consider the contrasts for this design and construct their variances and variance estimators, again for every case of the mixed effects model.

To determine the variance components and the ANOVA, we use a method explained by  \cite{Searle}. The design is completely randomized so that it makes sense to implement $F$ tests. The mathematical model is given by
\[y_{hijk}=m+R_h+A_i+e_{A_{hi}}+B_j+e_{B_{hj}}+AB_{ij}+e_{AB_{hij}}+C_k+AC_{ik}+BC_{jk}+ABC_{ijk}+e_t,\]
where $m$ is the general mean, $R_h$ is the $h$-th random block effect ($h=1,\ldots,r$), $A_i$ is the $i$-th horizontal strip effect ($i=1,\ldots,a$), $B_j$ is the $j$-th vertical strip effect ($j=1,\ldots,b$) and $C_k$ is the $k$-th effect on the intersection of $A$ and $B$ ($k=1,\ldots,c$). So $y_{hijk}$ represents the observation of the $i$-th level of $A$, the $j$-th level of $B$, the $k$-th level of $C$ on the block $h$. The errors $e_{A_{hi}}$, $e_{B_{hj}}$, $e_{AB_{hij}}$ and $e_{t_{hijk}}$ are normally distributed with mean zero and variance $\sigma_A^2$, $\sigma_B^2$, $\sigma_{AB}^2$ and $\sigma_t^2$, respectively. Since the blocks are random, we will assume $R\sim N(0,\sigma_R^2)$.

The analysis is done according to the scheme on Table \ref{tab0}, where $df$ stands for degrees of freedom and $SS$ stands for the sum of squares of the respective variation source.

\begin{table}[htpb]
\centering\small\caption{Sums of squares and Degrees of freedom}\label{tab0}
\begin{tabular}{|l|c|m{7cm}|}\hline
Source & $df$ &  \multicolumn{1}{|c|}{$SS$} \\ \hline
$R$ & $r-1$ & $abc\sum_{h=1}^r(\overline{y}_{h...}-\overline{y}_{....})^2$ \\[2mm]
$A$ & $a-1$ & $bcr\sum_{i=1}^a(\overline{y}_{.i..}-\overline{y}_{....})^2$\\[2mm]
$e_A$ & $(r-1)(a-1)$ & $bc\sum_{i=1}^a\sum_{h=1}^r(\overline{y}_{hi..}-\overline{y}_{h...}-\overline{y}_{.i..}+\overline{y}_{....})^2$ \\[2mm] \hline
$B$ & $b-1$ & $acr\sum_{j=1}^b(\overline{y}_{..j.}-\overline{y}_{....})^2$\\[2mm]
$e_B$ & $(r-1)(b-1)$ & $ac\sum_{j=1}^b\sum_{h=1}^r(\overline{y}_{h.j.}-\overline{y}_{h...}-\overline{y}_{..j.}+\overline{y}_{....})^2$ \\ [2mm]\hline
$AB$ & $(a-1)(b-1)$ & $cr\sum_{i=1}^a\sum_{j=1}^b(\overline{y}_{.ij.}-\overline{y}_{.i..}-\overline{y}_{..j.}+\overline{y}_{....})^2$\\[2mm]
$e_{AB}$ & $(a-1)(b-1)(r-1)$ & $c\sum_{i=1}^a\sum_{j=1}^b\sum_{h=1}^r(\overline{y}_{hij.}-	\overline{y}_{hi..}-\overline{y}_{h.j.}-\overline{y}_{.ij.}
+\overline{y}_{h...}+\overline{y}_{.i..}+\overline{y}_{..j.}-\overline{y}_{....})^2$\\[2mm] \hline
$C$ & $c-1$ & $abr\sum_{k=1}^c(\overline{y}_{...k}-\overline{y}_{....})^2$\\[2mm]
$AC$  & $(a-1)(c-1)$ & $br\sum_{i=1}^a\sum_{k=1}^c(\overline{y}_{.i.k}-\overline{y}_{.i..}-\overline{y}_{...k}+\overline{y}_{....})^2$\\[2mm]
$BC$  & $(b-1)(c-1)$ & $ar\sum_{j=1}^b\sum_{k=1}^c(\overline{y}_{..jk}-\overline{y}_{..j.}-\overline{y}_{...k}+\overline{y}_{....})^2$\\[2mm]
$ABC$ & $(a-1)(b-1)(c-1)$ & $r\sum_{i=1}^a\sum_{j=1}^b\sum_{k=1}^c(\overline{y}_{.ijk}-\overline{y}_{.i.k}-\overline{y}_{..jk}+\overline{y}_{...k}-\overline{y}_{.ij.}+\overline{y}_{.i..}+\overline{y}_{..j.}-\overline{y}_{....})^2$ \\[2mm]
$e_t$ & $ab(c-1)(r-1)$ & $\sum_{h=1}^r\sum_{i=1}^a\sum_{j=1}^b\sum_{k=1}^c(y_{hijk}-\overline{y}_{.ijk}-\overline{y}_{hij.}+\overline{y}_{.ij.})^2$\\ \hline
\end{tabular}
\end{table}

\section{Sums of squares and covariance matrices as Kronecker products}

Following the algorithm proposed by \cite {Moser}, the sums of squares and the covariance matrices can be computed using Kronecker products. Given the usefulness of this approach for computational purposes, we present those products here. Then the sums of squares can be expressed as:
\begin{align*}
	SS_R &=\boldsymbol{Y'}[(\boldsymbol{I}_{r}-r^{-1}\boldsymbol{J}_{r})\otimes 							a^{-1}\boldsymbol{J}_{a}\otimes b^{-1}\boldsymbol{J}_{b}\otimes c^{-1}\boldsymbol{J}_{c}]			\boldsymbol{Y},\\
	SS_A  &=\boldsymbol{Y'}[r^{-1}\boldsymbol{J}_{r}\otimes(\boldsymbol{I}_{a}-							a^{-1}\boldsymbol{J}_{a})\otimes b^{-1}\boldsymbol{J}_{b}\otimes            							c^{-1}\boldsymbol{J}_{c}]\boldsymbol{Y},\\
	SS_{e_A}&=\boldsymbol{Y'}[(\boldsymbol{I}_{r}-r^{-1}\boldsymbol{J}_{r})\otimes               					(\boldsymbol{I}_{a}-a^{-1}\boldsymbol{J}_{a})\otimes b^{-1}\boldsymbol{J}_{b}\otimes 				c^{-1}\boldsymbol{J}_{c}]\boldsymbol{Y},\\
	SS_B &=\boldsymbol{Y'}[r^{-1}\boldsymbol{J}_{r}\otimes a^{-1}\boldsymbol{J}_{a}\otimes                                                      		(\boldsymbol{I}_{b}-b^{-1}\boldsymbol{J}_{b})\otimes c^{-1}\boldsymbol{J}_{c}]\boldsymbol{Y},\\
	SS_{e_B} &=\boldsymbol{Y'}[(\boldsymbol{I}_{r}-r^{-1}\boldsymbol{J}_{r})\otimes               				a^{-1}\boldsymbol{J}_{a}\otimes (\boldsymbol{I}_{b}-b^{-1}\boldsymbol{J}_{b})\otimes                                 		c^{-1}\boldsymbol{J}_{c}]\boldsymbol{Y},\\
	SS_{AB} &=\boldsymbol{Y'}[{r}^{-1}\boldsymbol{J}_{r}\otimes(\boldsymbol{I}_{a}-						a^{-1}\boldsymbol{J}_{a})\otimes(\boldsymbol{I}_{b}-b^{-1}\boldsymbol{J}_{b})\otimes                                 		c^{-1}\boldsymbol{J}_{c}]\boldsymbol{Y},\\
	SS_{e_{AB}} &=\boldsymbol{Y'}[(\boldsymbol{I}_{r}-r^{-1}\boldsymbol{J}_{r})\otimes               				(\boldsymbol{I}_{a}-a^{-1}\boldsymbol{J}_{a})\otimes(\boldsymbol{I}_{b}-							b^{-1}\boldsymbol{J}_{b})\otimes c^{-1}\boldsymbol{J}_{c}]\boldsymbol{Y},\\
	SS_C &=\boldsymbol{Y'}[r^{-1}\boldsymbol{J}_{r}\otimes a^{-1}\boldsymbol{J}_{a}\otimes 				b^{-1}\boldsymbol{J}_{b}\otimes(\boldsymbol{I}_{c}-c^{-1}\boldsymbol{J}_{c})]\boldsymbol{Y},\\
	SS_{AC} &=\boldsymbol{Y'}[r^{-1}\boldsymbol{J}_{r}\otimes(\boldsymbol{I}_{a}-						a^{-1}\boldsymbol{J}_{a})\otimes b^{-1}\boldsymbol{J}_{b}\otimes                                                            		(\boldsymbol{I}_{c}-c^{-1}\boldsymbol{J}_{c})]\boldsymbol{Y},\\
	SS_{BC} &=\boldsymbol{Y'}[r^{-1}\boldsymbol{J}_{r}\otimes a^{-1}\boldsymbol{J}_{a}\otimes                                                      		(\boldsymbol{I}_{b}-b^{-1}\boldsymbol{J}_{b})\otimes (\boldsymbol{I}_{c}-						c^{-1}\boldsymbol{J}_{c})]\boldsymbol{Y},\\
	SS_{ABC} &=\boldsymbol{Y'}[r^{-1}\boldsymbol{J}_{r}\otimes (\boldsymbol{I}_{a}-						a^{-1}\boldsymbol{J}_{a})\otimes (\boldsymbol{I}_{b}-b^{-1}\boldsymbol{J}_{b})\otimes                                 		(\boldsymbol{I}_{c}-c^{-1}\boldsymbol{J}_{c})]\boldsymbol{Y},\\
	SS_{e_t}&=\boldsymbol{Y'}[\boldsymbol{I}_{r}\otimes (\boldsymbol{I}_{a}-a^{-1}\boldsymbol{J}_{a})		\otimes(\boldsymbol{I}_{b}-b^{-1}\boldsymbol{J}_{b})\otimes\boldsymbol{I}_{c}]\boldsymbol{Y},
\end{align*}
where $\boldsymbol J_j$ is a squared matrix of ones and size $j$, and $\boldsymbol I_i$ is an identity matrix of size $i$.

The covariance matrix when all effects are fixed is:
\begin{align*}
	V(\boldsymbol{Y})&=\sigma_R^2(\boldsymbol{I}_{r}\otimes \boldsymbol{J}_{a}\otimes 					\boldsymbol{J}_{b}\otimes \boldsymbol{J}_{c})+
		\sigma_{e_A}^2(\boldsymbol{I}_{r}\otimes \boldsymbol{I}_{a}\otimes \boldsymbol{J}_{b}\otimes 		\boldsymbol{J}_{c})+\\
	&\sigma_{e_B}^2(\boldsymbol{I}_{r}\otimes \boldsymbol{J}_{a}\otimes \boldsymbol{I}_{b}\otimes 			\boldsymbol{J}_{c})+
		\sigma_{e_{AB}}^2(\boldsymbol{I}_{r}\otimes \boldsymbol{I}_{a}\otimes \boldsymbol{I}_{b}			\otimes \boldsymbol{J}_{c})+\\
	&\sigma_{e_t}^2(\boldsymbol{I}_{r}\otimes \boldsymbol{I}_{a}\otimes \boldsymbol{I}_{b}			\otimes \boldsymbol{I}_{c}).
\end{align*}

The covariance matrix when all effects are random is:
\begin{align*}
	V(\boldsymbol{Y})&=\sigma_R^2(\boldsymbol{I}_{r}\otimes \boldsymbol{J}_{a}\otimes 					\boldsymbol{J}_{b}\otimes \boldsymbol{J}_{c})+
		\sigma_A^2(\boldsymbol{J}_{r}\otimes \boldsymbol{I}_{a}\otimes \boldsymbol{J}_{b}\otimes 			\boldsymbol{J}_{c})+\\
	&\sigma_{e_A}^2(\boldsymbol{I}_{r}\otimes \boldsymbol{I}_{a}\otimes \boldsymbol{J}_{b}\otimes 			\boldsymbol{J}_{c})+
		\sigma_B^2(\boldsymbol{J}_{r}\otimes \boldsymbol{J}_{a}\otimes \boldsymbol{I}_{b}\otimes 			\boldsymbol{J}_{c})+\\
	&\sigma_{e_B}^2(\boldsymbol{I}_{r}\otimes \boldsymbol{J}_{a}\otimes \boldsymbol{I}_{b}\otimes 			\boldsymbol{J}_{c})+
		\sigma_{AB}^2(\boldsymbol{J}_{r}\otimes \boldsymbol{I}_{a}\otimes \boldsymbol{I}_{b}\otimes 		\boldsymbol{J}_{c})+\\
	&\sigma_{e_{AB}}^2(\boldsymbol{I}_{r}\otimes \boldsymbol{I}_{a}\otimes \boldsymbol{I}_{b}\otimes 		\boldsymbol{J}_{c})+
		\sigma_C^2(\boldsymbol{J}_{r}\otimes \boldsymbol{J}_{a}\otimes \boldsymbol{J}_{b}\otimes 			\boldsymbol{I}_{c})+\\
	&\sigma_{AC}^2(\boldsymbol{J}_{r}\otimes \boldsymbol{I}_{a}\otimes \boldsymbol{J}_{b}\otimes 			\boldsymbol{I}_{c})+
		\sigma_{BC}^2(\boldsymbol{J}_{r}\otimes \boldsymbol{J}_{a}\otimes \boldsymbol{I}_{b}\otimes 		\boldsymbol{I}_{c})+\\
	&\sigma_{ABC}^2(\boldsymbol{J}_{r}\otimes \boldsymbol{I}_{a}\otimes \boldsymbol{I}_{b}\otimes 			\boldsymbol{I}_{c})+
		\sigma_{e_t}^2(\boldsymbol{I}_{r}\otimes \boldsymbol{I}_{a}\otimes \boldsymbol{I}_{b}				\otimes \boldsymbol{I}_{c}).
\end{align*}

The covariance matrix when only $A$ has fixed effects is:
\begin{align*} 
	V(\boldsymbol{Y})&=\sigma_R^2(\boldsymbol{I}_{r}\otimes \boldsymbol{J}_{a}\otimes 					\boldsymbol{J}_{b}\otimes \boldsymbol{J}_{c})+
		\sigma_{e_A}^2(\boldsymbol{I}_{r}\otimes \boldsymbol{I}_{a}\otimes \boldsymbol{J}_{b}\otimes 		\boldsymbol{J}_{c})+\\
	&\sigma_B^2(\boldsymbol{J}_{r}\otimes \boldsymbol{J}_{a}\otimes \boldsymbol{I}_{b}\otimes 				\boldsymbol{J}_{c})+
		\sigma_{e_B}^2(\boldsymbol{I}_{r}\otimes \boldsymbol{J}_{a}\otimes \boldsymbol{I}_{b}\otimes 		\boldsymbol{J}_{c})+\\
	&\sigma_{AB}^2(\boldsymbol{J}_{r}\otimes \boldsymbol{I}_{a}\otimes \boldsymbol{I}_{b}\otimes 			\boldsymbol{J}_{c})+
		\sigma_{e_{AB}}^2(\boldsymbol{I}_{r}\otimes \boldsymbol{I}_{a}\otimes \boldsymbol{I}_{b}			\otimes \boldsymbol{J}_{c})+\\
	&\sigma_C^2(\boldsymbol{J}_{r}\otimes \boldsymbol{J}_{a}\otimes \boldsymbol{J}_{b}\otimes 			\boldsymbol{I}_{c})+
		\sigma_{AC}^2(\boldsymbol{J}_{r}\otimes \boldsymbol{I}_{a}\otimes \boldsymbol{J}_{b}\otimes 		\boldsymbol{I}_{c})+\\
	&\sigma_{BC}^2(\boldsymbol{J}_{r}\otimes \boldsymbol{J}_{a}\otimes \boldsymbol{I}_{b}\otimes 			\boldsymbol{I}_{c})+
		\sigma_{ABC}^2(\boldsymbol{J}_{r}\otimes \boldsymbol{I}_{a}\otimes \boldsymbol{I}_{b}\otimes 		\boldsymbol{I}_{c})+\\
	&\sigma_{e_t}^2(\boldsymbol{I}_{r}\otimes \boldsymbol{I}_{a}\otimes \boldsymbol{I}_{b}\otimes 			\boldsymbol{I}_{c}).
\end{align*}

The covariance matrix when only $B$ has fixed effects is:
\begin{align*} 
	V(\boldsymbol{Y})&=\sigma_R^2(\boldsymbol{I}_{r}\otimes \boldsymbol{J}_{a}\otimes 					\boldsymbol{J}_{b}\otimes \boldsymbol{J}_{c})+
		\sigma_A^2(\boldsymbol{J}_{r}\otimes \boldsymbol{I}_{a}\otimes \boldsymbol{J}_{b}\otimes 			\boldsymbol{J}_{c})+\\
	&\sigma_{e_A}^2(\boldsymbol{I}_{r}\otimes \boldsymbol{I}_{a}\otimes \boldsymbol{J}_{b}\otimes 			\boldsymbol{J}_{c})+
		\sigma_{e_B}^2(\boldsymbol{I}_{r}\otimes \boldsymbol{J}_{a}\otimes \boldsymbol{I}_{b}\otimes 		\boldsymbol{J}_{c})+\\
	&\sigma_{AB}^2(\boldsymbol{J}_{r}\otimes \boldsymbol{I}_{a}\otimes \boldsymbol{I}_{b}\otimes 			\boldsymbol{J}_{c})+
		\sigma_{e_{AB}}^2(\boldsymbol{I}_{r}\otimes \boldsymbol{I}_{a}\otimes \boldsymbol{I}_{b}			\otimes \boldsymbol{J}_{c})+\\
	&\sigma_C^2(\boldsymbol{J}_{r}\otimes \boldsymbol{J}_{a}\otimes \boldsymbol{J}_{b}\otimes 			\boldsymbol{I}_{c})+
		\sigma_{AC}^2(\boldsymbol{J}_{r}\otimes \boldsymbol{I}_{a}\otimes \boldsymbol{J}_{b}\otimes 		\boldsymbol{I}_{c})+\\
	&\sigma_{BC}^2(\boldsymbol{J}_{r}\otimes \boldsymbol{J}_{a}\otimes \boldsymbol{I}_{b}\otimes 			\boldsymbol{I}_{c})+
		\sigma_{ABC}^2(\boldsymbol{J}_{r}\otimes \boldsymbol{I}_{a}\otimes \boldsymbol{I}_{b}\otimes 		\boldsymbol{I}_{c})+\\
	&\sigma_{e_t}^2(\boldsymbol{I}_{r}\otimes \boldsymbol{I}_{a}\otimes \boldsymbol{I}_{b}\otimes 		\boldsymbol{I}_{c}).
\end{align*}

The covariance matrix when only $C$ has fixed effects is:
\begin{align*} 
	V(\boldsymbol{Y})&=\sigma_R^2(\boldsymbol{I}_{r}\otimes \boldsymbol{J}_{a}\otimes 					\boldsymbol{J}_{b}\otimes \boldsymbol{J}_{c})+
		\sigma_A^2(\boldsymbol{J}_{r}\otimes \boldsymbol{I}_{a}\otimes \boldsymbol{J}_{b}\otimes 			\boldsymbol{J}_{c})+\\
	&\sigma_{e_A}^2(\boldsymbol{I}_{r}\otimes \boldsymbol{I}_{a}\otimes \boldsymbol{J}_{b}\otimes 			\boldsymbol{J}_{c})+
		\sigma_B^2(\boldsymbol{J}_{r}\otimes \boldsymbol{J}_{a}\otimes \boldsymbol{I}_{b}\otimes 			\boldsymbol{J}_{c})+\\
	&\sigma_{e_B}^2(\boldsymbol{I}_{r}\otimes \boldsymbol{J}_{a}\otimes \boldsymbol{I}_{b}\otimes 			\boldsymbol{J}_{c})+
		\sigma_{AB}^2(\boldsymbol{J}_{r}\otimes \boldsymbol{I}_{a}\otimes \boldsymbol{I}_{b}\otimes 		\boldsymbol{J}_{c})+\\
	&\sigma_{e_{AB}}^2(\boldsymbol{I}_{r}\otimes \boldsymbol{I}_{a}\otimes \boldsymbol{I}_{b}\otimes 		\boldsymbol{J}_{c})+
		\sigma_{AC}^2(\boldsymbol{J}_{r}\otimes \boldsymbol{I}_{a}\otimes \boldsymbol{J}_{b}\otimes 		\boldsymbol{I}_{c})+\\
	&\sigma_{BC}^2(\boldsymbol{J}_{r}\otimes \boldsymbol{J}_{a}\otimes \boldsymbol{I}_{b}\otimes 			\boldsymbol{I}_{c})+
		\sigma_{ABC}^2(\boldsymbol{J}_{r}\otimes \boldsymbol{I}_{a}\otimes \boldsymbol{I}_{b}\otimes 		\boldsymbol{I}_{c})+\\
	&\sigma_{e_t}^2(\boldsymbol{I}_{r}\otimes \boldsymbol{I}_{a}\otimes \boldsymbol{I}_{b}\otimes 			\boldsymbol{I}_{c}).
\end{align*}

The covariance matrix when only $A$ has random effects is:
\begin{align*} 
	V(\boldsymbol{Y})&=\sigma_R^2(\boldsymbol{I}_{r}\otimes \boldsymbol{J}_{a}\otimes 					\boldsymbol{J}_{b}\otimes \boldsymbol{J}_{c})+
		\sigma_A^2(\boldsymbol{J}_{r}\otimes \boldsymbol{I}_{a}\otimes \boldsymbol{J}_{b}\otimes 			\boldsymbol{J}_{c})+\\
	&\sigma_{e_A}^2(\boldsymbol{I}_{r}\otimes \boldsymbol{I}_{a}\otimes \boldsymbol{J}_{b}\otimes 			\boldsymbol{J}_{c})+
		\sigma_{e_B}^2(\boldsymbol{I}_{r}\otimes \boldsymbol{J}_{a}\otimes \boldsymbol{I}_{b}\otimes 		\boldsymbol{J}_{c})+\\
	&\sigma_{AB}^2(\boldsymbol{J}_{r}\otimes \boldsymbol{I}_{a}\otimes \boldsymbol{I}_{b}\otimes 			\boldsymbol{J}_{c})+
		\sigma_{e_{AB}}^2(\boldsymbol{I}_{r}\otimes \boldsymbol{I}_{a}\otimes \boldsymbol{I}_{b}			\otimes \boldsymbol{J}_{c})+\\
	&\sigma_{AC}^2(\boldsymbol{J}_{r}\otimes \boldsymbol{I}_{a}\otimes \boldsymbol{J}_{b}\otimes 			\boldsymbol{I}_{c})+
		\sigma_{ABC}^2(\boldsymbol{J}_{r}\otimes \boldsymbol{I}_{a}\otimes \boldsymbol{I}_{b}\otimes 		\boldsymbol{I}_{c})+\\
	&\sigma_{e_t}^2(\boldsymbol{I}_{r}\otimes \boldsymbol{I}_{a}\otimes \boldsymbol{I}_{b}					\otimes \boldsymbol{I}_{c}).
\end{align*}

The covariance matrix when only $B$ has random effects is:
\begin{align*} 
	V(\boldsymbol{Y})&=\sigma_R^2(\boldsymbol{I}_{r}\otimes \boldsymbol{J}_{a}\otimes 					\boldsymbol{J}_{b}\otimes \boldsymbol{J}_{c})+
		\sigma_{e_A}^2(\boldsymbol{I}_{r}\otimes \boldsymbol{I}_{a}\otimes \boldsymbol{J}_{b}\otimes 		\boldsymbol{J}_{c})+\\
	&\sigma_B^2(\boldsymbol{J}_{r}\otimes \boldsymbol{J}_{a}\otimes \boldsymbol{I}_{b}\otimes 				\boldsymbol{J}_{c})+
		\sigma_{e_B}^2(\boldsymbol{I}_{r}\otimes \boldsymbol{J}_{a}\otimes \boldsymbol{I}_{b}\otimes 		\boldsymbol{J}_{c})+\\
	&\sigma_{AB}^2(\boldsymbol{J}_{r}\otimes \boldsymbol{I}_{a}\otimes \boldsymbol{I}_{b}\otimes 			\boldsymbol{J}_{c})+
		\sigma_{e_{AB}}^2(\boldsymbol{I}_{r}\otimes \boldsymbol{I}_{a}\otimes \boldsymbol{I}_{b}			\otimes \boldsymbol{J}_{c})+\\
	&\sigma_{BC}^2(\boldsymbol{J}_{r}\otimes \boldsymbol{J}_{a}\otimes \boldsymbol{I}_{b}\otimes 			\boldsymbol{I}_{c})+
		\sigma_{ABC}^2(\boldsymbol{J}_{r}\otimes \boldsymbol{I}_{a}\otimes \boldsymbol{I}_{b}\otimes 		\boldsymbol{I}_{c})+\\
	&\sigma_{e_t}^2(\boldsymbol{I}_{r}\otimes \boldsymbol{I}_{a}\otimes \boldsymbol{I}_{b}\otimes 			\boldsymbol{I}_{c}).
\end{align*}

The covariance matrix when only $C$ has random effects is:
\begin{align*} 
	V(\boldsymbol{Y})&=\sigma_R^2(\boldsymbol{I}_{r}\otimes \boldsymbol{J}_{a}\otimes 					\boldsymbol{J}_{b}\otimes \boldsymbol{J}_{c})+
		\sigma_{e_A}^2(\boldsymbol{I}_{r}\otimes \boldsymbol{I}_{a}\otimes \boldsymbol{J}_{b}\otimes 		\boldsymbol{J}_{c})+\\
	&\sigma_{e_B}^2(\boldsymbol{I}_{r}\otimes \boldsymbol{J}_{a}\otimes \boldsymbol{I}_{b}\otimes 			\boldsymbol{J}_{c})+
		\sigma_{e_{AB}}^2(\boldsymbol{I}_{r}\otimes \boldsymbol{I}_{a}\otimes \boldsymbol{I}_{b}			\otimes \boldsymbol{J}_{c})+\\
	&\sigma_C^2(\boldsymbol{J}_{r}\otimes \boldsymbol{J}_{a}\otimes \boldsymbol{J}_{b}\otimes 			\boldsymbol{I}_{c})+
		\sigma_{AC}^2(\boldsymbol{J}_{r}\otimes \boldsymbol{I}_{a}\otimes \boldsymbol{J}_{b}\otimes 		\boldsymbol{I}_{c})+\\
	&\sigma_{BC}^2(\boldsymbol{J}_{r}\otimes \boldsymbol{J}_{a}\otimes \boldsymbol{I}_{b}\otimes 			\boldsymbol{I}_{c})+
		\sigma_{ABC}^2(\boldsymbol{J}_{r}\otimes \boldsymbol{I}_{a}\otimes \boldsymbol{I}_{b}\otimes 		\boldsymbol{I}_{c})+\\
	&\sigma_{e_t}^2(\boldsymbol{I}_{r}\otimes \boldsymbol{I}_{a}\otimes \boldsymbol{I}_{b}					\otimes \boldsymbol{I}_{c}).
\end{align*}

\section{Expected mean squares}

To illustrate how to obtain the expected mean squares $E(MS)$ we will show the process for the random blocks $R$ (For the remaining cases, since the procedure is similar, we will only present the final value without the respective development):   First, take $SS_R$ in Table \ref{tab0} and calculate its expected value:
\begin{align*}
	E(SS_R)=abc\sum_{h=1}^{r}E(R_h-\overline{R}_{.}+\overline{e}_{A_{h.}}-\overline{e}_{A_{..}}+			\overline{e}_{B_{h.}}-\overline{e}_{B_{..}}+\overline{e}_{AB_{h..}}-\overline{e}_{AB_{...}}+				\overline{e}_{t_{h...}}-\overline{e}_{t_{....}})^2,
\end{align*}
since the product of errors and factors is always zero under expectation, $E(SS_R)$ equals
\begin{align*}
	abc\sum_{h=1}^{r}E(R_h-\overline{R}_{.})^2+abc\sum_{h=1}^{r}E(\overline{e}_{A_{h.}}-					\overline{e}_{A_{..}}+\overline{e}_{B_{h.}}-\overline{e}_{B_{..}}+\overline{e}_{AB_{h..}}-				\overline{e}_{AB_{...}}+\overline{e}_{t_{h...}}-\overline{e}_{t_{....}})^2,
\end{align*}
and since the errors are independent between themselves,
\begin{align*}
	E(SS_R)&=abc\sum_{h=1}^{r}E(R_h-\overline{R}_{.})^2+abc\sum_{h=1}^{r}E(\overline{e}_{A_{h.}}-		\overline{e}_{A_{..}})^2+abc\sum_{h=1}^{r}E(\overline{e}_{B_{h.}}-\overline{e}_{B_{..}})^2\\
	&+abc\sum_{h=1}^{r}E(\overline{e}_{AB_{h..}}-\overline{e}_{AB_{...}})^2+abc							\sum_{h=1}^{r}E(\overline{e}_{t_{h...}}-\overline{e}_{t_{....}})^2.
\end{align*}
Therefore, taking into account that $\sigma_{e}^2=E(e^2)-E^2(e)$, and that $E(e)=0$ for every error in the model,
\begin{align*}
	E(SS_R)   &=abc\sum_{h=1}^{r}E(R_h-\overline{R}_{.})^2+abc\frac{(r-1)\sigma_{e_A}^2}{a}\\
		&+abc\frac{(r-1)\sigma_{e_B}^2}{b}+abc\frac{(r-1)\sigma_{e_{AB}}^2}{ab}+abc						\frac{(r-1)\sigma_{e_t}^2}{abc}.
\end{align*}
Then, taking $E(SS_R)$ and dividing it by its $df$, we obtain the expected mean square for fixed blocks:
\begin{align}\label{MS}
	E(MS_R)   =\frac{abc}{r-1}\sum_{h=1}^{r}E(R_h-\overline{R}_{.})^2+bc\sigma_{e_A}^2+ac				\sigma_{e_B}^2+c\sigma_{e_{AB}}^2+\sigma_{e_t}^2.
\end{align}

Now, for the more interesting case of random blocks, we get:
\begin{align*}
	E(MS_R)   &=abc\sigma_R^2+bc\sigma_{e_A}^2+ac\sigma_{e_B}^2+c\sigma_{e_AB}^2+				\sigma_{e_t}^2.
\end{align*}

Note that $E(MS_R)$ will remain unchanged regardless the model we are considering. This is also true for the expectation of the mean square of each error involved. So we mention these here and will omit them in the particular description of the $E(MS)$'s for each model:
\begin{align*}
	E(MS_{e_A})&=bc\sigma_{e_A}^2+c\sigma_{e_{AB}}^2+\sigma_{e_t}^2,\\
	E(MS_{e_B})&=ac\sigma_{e_B}^2+c\sigma_{e_{AB}}^2+\sigma_{e_t}^2,\\
	E(MS_{e_{AB}})  &=c\sigma_{e_{AB}}^2+\sigma_{e_t}^2,\\
	E(MS_{e_t})&=\sigma_{e_t}^2.
\end{align*}

Finally, note also that every interaction involving a random effect will be random. So in the following subsections, to avoid confusion, we present explicitly all the $E(MS)$'s for every model.

\subsection{Expected mean squares for the fixed effects model}\label{fixed}

When the effects are fixed (constant), by definition it is sufficient to suppress the expectation operator of the mean squares considered. Thus we get:
\begin{align*}
	E(MS_A)&=\frac{bcr}{a-1}\sum_{i=1}^{a}(A_i-\overline{A}_{.}+\overline{AB}_{i.}-\overline{AB}_{..}+				\overline{AC}_{i.}-\overline{AC}_{..}+\overline{ABC}_{i..}-\overline{ABC}_{...})^2\\
		&+bc\sigma_{e_A}^2+c\sigma_{e_{AB}}^{2}+\sigma_{e_t}^2,\\
	E(MS_B)&=\frac{acr}{b-1}\sum_{j=1}^{b}(B_j-\overline{B}_{.}+\overline{AB}_{.j}-\overline{AB}_{..}+				\overline{BC}_{j.}-\overline{BC}_{..}+\overline{ABC}_{.j.}-\overline{ABC}_{...})^2\\
             	&+ac\sigma_{e_B}^2+c\sigma_{e_{AB}}^{2}+\sigma_{e_t}^2,\\
	E(MS_{AB})&=\frac{cr}{(a-1)(b-1)}\sum_{i=1}^{a}\sum_{j=1}^{b}(AB_{ij}-\overline{AB}_{.j}-					\overline{AB}_{i.}+\overline{AB}_{..}\\
             	&+\overline{ABC}_{ij.}-\overline{ABC}_{.j.}-\overline{ABC}_{i..}+\overline{ABC}_{...})^2+c					\sigma_{e_{AB}}^2+\sigma_{e_t}^2,\\
	E(MS_C)&=\frac{abr}{c-1}\sum_{k=1}^{c}(C_k-\overline{C}_{.}+\overline{AC}_{.k}-\overline{AC}_{..}+				\overline{BC}_{.k}-\overline{BC}_{..}+\overline{ABC}_{..k}-\overline{ABC}_{...})^2+					\sigma_{e_t}^2,\\
	E(MS_{AC})&=\frac{br}{(a-1)(c-1)}\sum_{i=1}^{a}\sum_{k=1}^{c}(AC_{ik}-\overline{AC}_{.k}-					\overline{AC}_{i.}+\overline{AC}_{..}\\
             	&+\overline{ABC}_{i.k}-\overline{ABC}_{..k}-\overline{ABC}_{i..}+\overline{ABC}_{...})^2+					\sigma_{e_t}^2,\\
	E(MS_{BC})&=\frac{ar}{(b-1)(c-1)}\sum_{j=1}^{b}\sum_{k=1}^{c}(BC_{jk}-\overline{BC}_{.k}-					\overline{BC}_{j.}+\overline{BC}_{..}\\
             	&+\overline{ABC}_{.jk}-\overline{ABC}_{..k}-\overline{ABC}_{.j.}+\overline{ABC}_{...})^2+					\sigma_{e_t}^2,\\
	E(MS_{ABC})&=\frac{r}{(a-1)(b-1)(c-1)}\sum_{i=1}^{a}\sum_{j=1}^{b}\sum_{k=1}^{c}(ABC_{ijk}-					\overline{ABC}_{.jk}\\
             	&-\overline{ABC}_{i.k}+\overline{ABC}_{..k}-\overline{ABC}_{ij.}+\overline{ABC}_{.j.}					+\overline{ABC}_{i..}-\overline{ABC}_{...})^2+\sigma_{e_t}^2.
\end{align*}

\subsection{Expected mean squares for the random effects model}\label{random}

When a factor, say $A$, has random effects, we will assume that the effects of A have distribution $N(0,\sigma_A^2)$. Then for the random effects model, the effects of $A$, $B$ and $C$ will be random, independent, and normally distributed with mean 0 and variance $\sigma_A^2$, $\sigma_B^2$ and $\sigma_C^2$, respectively.  The interactions $AB$, $AC$, $BC$ and $ABC$ will have normal distribution with mean 0 and variance $\sigma_{AB}^2$, $\sigma_{AC}^2$, $\sigma_{BC}^2$ and $\sigma_{ABC}^2$, respectively. We also assume that the effects are independent between them. So we get:

\begin{align*}
	E(MS_A)&=bcr\sigma_{A}^2+bc\sigma_{e_A}^2+cr\sigma_{AB}^2+c\sigma_{e_{AB}}^2,+br				\sigma_{AC}^2+r\sigma_{ABC}^2+\sigma_{e_t}^2,\\
	E(MS_B)&=acr\sigma_{B}^2+ac\sigma_{e_B}^2+cr\sigma_{AB}^2+c\sigma_{e_{AB}}^2,+ar				\sigma_{BC}^2+r\sigma_{ABC}^2+\sigma_{e_t}^2,\\
	E(MS_{AB})&=cr\sigma_{AB}^2+c\sigma_{e_{AB}}^2+r\sigma_{ABC}^2+\sigma_{e_t}^2,\\
	E(MSC)&=abr\sigma_{C}^2+br\sigma_{AC}^2+ar\sigma_{BC}^2+r\sigma_{ABC}^2+					\sigma_{e_t}^2,\\
	E(MS_{AC})&=br\sigma_{AC}^2+r\sigma_{ABC}^2+\sigma_{e_t}^2,\\
	E(MS_{BC})&=ar\sigma_{BC}^2+r\sigma_{ABC}^2+\sigma_{e_t}^2,\\
	E(MS_{ABC})&=r\sigma_{ABC}^2+\sigma_{e_t}^2.
\end{align*}

\subsection{Expected mean squares when only $A$ is fixed}\label{Afixed}

In this case, $B$ and $C$ will be random with variances $\sigma_B^2$ and $\sigma_C^2$, respectively. Also, $AB$, $AC$, $BC$ and $ABC$ will be random with variances $\sigma_{AB}^2$, $\sigma_{AC}^2$, $\sigma_{BC}^2$ and $\sigma_{ABC}^2$, respectively. So we obtain:

\begin{align*}
	E(MS_A)&=\frac{bcr}{a-1}\sum_{i=1}^{a}(A_i-\overline{A}_{.})^2+bc\sigma_{e_A}^2+cr\sigma_{AB}^2		+c\sigma_{e_{AB}}^2+br\sigma_{AC}^2+r\sigma_{ABC}^2+\sigma_{e_t}^2,\\
	E(MS_B)&=acr\sigma_{B}^2+ac\sigma_{e_B}^2+cr\sigma_{AB}^2+c\sigma_{e_{AB}}^2
		+ar\sigma_{BC}^2+r\sigma_{ABC}^2+\sigma_{e_t}^2,\\
	E(MS_{AB})&=cr\sigma_{AB}^2+c\sigma_{e_{AB}}^2+r\sigma_{ABC}^2+\sigma_{e_t}^2,\\
	E(MS_C)&=abr\sigma_{C}^2+br\sigma_{AC}^2+ar\sigma_{BC}^2+r\sigma_{ABC}^2+\sigma_{e_t}^2
		,\\
	E(MS_{AC})&=br\sigma_{AC}^2+r\sigma_{ABC}^2+\sigma_{e_t}^2,\\
	E(MS_{BC})&=ar\sigma_{BC}^2+r\sigma_{ABC}^2+\sigma_{e_t}^2,\\
	E(MS_{ABC})&=r\sigma_{ABC}^2+\sigma_{e_t}^2.    
\end{align*}

\subsection{Expected mean squares when only $B$ is fixed}\label{Bfixed}

Here we have that $A$ and $C$ will be random with variances $\sigma_A^2$ and $\sigma_C^2$, respectively. Also, $AB$, $AC$, $BC$ and $ABC$ will be random with variances $\sigma_{AB}^2$, $\sigma_{AC}^2$, $\sigma_{BC}^2$ and $\sigma_{ABC}^2$, respectively. So we obtain:

\begin{align*}
	E(MS_A)&=bcr\sigma_{A}^2+bc\sigma_{e_A}^2+cr\sigma_{AB}^2+c\sigma_{e_{AB}}^2
		+br\sigma_{AC}^2+r\sigma_{ABC}^2+\sigma_{e_t}^2,\\
	E(MS_B)&=\frac{acr}{b-1}\sum_{j=1}^{b}(B_j-\overline{B}_{.})^2+ac\sigma_{e_B}^2+cr\sigma_{AB}^2
		+c\sigma_{e_{AB}}^2+ar\sigma_{BC}^2+r\sigma_{ABC}^2+\sigma_{e_t}^2,\\
	E(MS_{AB})&=cr\sigma_{AB}^2+c\sigma_{e_{AB}}^2+r\sigma_{ABC}^2+\sigma_{e_t}^2,\\
	E(MS_C)&=abr\sigma_{C}^2+br\sigma_{AC}^2+ar\sigma_{BC}^2+r\sigma_{ABC}^2+\sigma_{e_t}^2
		,\\
	E(MS_{AC})&=br\sigma_{AC}^2+r\sigma_{ABC}^2+\sigma_{e_t}^2,\\
	E(MS_{BC})&=ar\sigma_{BC}^2+r\sigma_{ABC}^2+\sigma_{e_t}^2,\\
	E(MS_{ABC})&=r\sigma_{ABC}^2+\sigma_{e_t}^2.
\end{align*}

\subsection{Expected mean squares when only $C$ is fixed}\label{Cfixed}

In this case, $A$ and $B$ are random with variances $\sigma_A^2$ and $\sigma_B^2$, respectively. Also, $AB$, $AC$, $BC$ and $ABC$ will be random with variances $\sigma_{AB}^2$, $\sigma_{AC}^2$, $\sigma_{BC}^2$ and $\sigma_{ABC}^2$, respectively. So we obtain:

\begin{align*}
	E(MS_A)&=bcr\sigma_{A}^2+bc\sigma_{e_A}^2+cr\sigma_{AB}^2+c\sigma_{e_{AB}}^2
		+br\sigma_{AC}^2+r\sigma_{ABC}^2+\sigma_{e_t}^2,\\
	E(MS_B)&=acr\sigma_{B}^2+ac\sigma_{e_B}^2+cr\sigma_{AB}^2+c\sigma_{e_{AB}}^2
		+ar\sigma_{BC}^2+r\sigma_{ABC}^2+\sigma_{e_t}^2,\\
	E(MS_{AB})&=cr\sigma_{AB}^2+c\sigma_{e_{AB}}^2+r\sigma_{ABC}^2+\sigma_{e_t}^2,\\
	E(MS_C)&=\frac{abr}{c-1}\sum_{k=1}^{c}(C_k-\overline{C}_{.})^2+br\sigma_{AC}^2+ar\sigma_{BC}^2
		+r\sigma_{ABC}^2+\sigma_{e_t}^2,\\
	E(MS_{AC})&=br\sigma_{AC}^2+r\sigma_{ABC}^2+\sigma_{e_t}^2,\\
	E(MS_{BC})&=ar\sigma_{BC}^2+r\sigma_{ABC}^2+\sigma_{e_t}^2,\\
	E(MS_{ABC})&=r\sigma_{ABC}^2+\sigma_{e_t}^2.
\end{align*}

\subsection{Expected mean squares when only $A$ is random}\label{Arandom}

In this case, $A$, $AB$, $AC$ and $ABC$ are random with variance $\sigma_A^2$, $\sigma_{AB}^2$, $\sigma_{AC}^2$ and $\sigma_{ABC}^2$, respectively. Thus, we get:

\begin{align*}
	E(MS_A) &=bcr\sigma_{A}^2+bc\sigma_{e_A}^2+cr\sigma_{AB}^2+c\sigma_{e_{AB}}^2
		+br\sigma_{AC}^2+r\sigma_{ABC}^2+\sigma_{e_t}^2,\\
	E(MS_B)&=\frac{acr}{b-1}\sum_{j=1}^{b}(B_j-\overline{B}_{.}+\overline{BC}_{j.}-\overline{BC}_{..})^2
             	+ac\sigma_{e_B}^2+cr\sigma_{AB}^2+c\sigma_{e_{AB}}^2+r\sigma_{ABC}^2+\sigma_{e_t}^2,\\
	E(MS_{AB})&=cr\sigma_{AB}^2+c\sigma_{e_{AB}}^2+r\sigma_{ABC}^2+\sigma_{e_t}^2,\\
	E(MS_C)&=\frac{abr}{c-1}\sum_{k=1}^{c}(C_k-\overline{C}_{.}+\overline{BC}_{.k}-						\overline{BC}_{..})^2+br\sigma_{AC}^2+r\sigma_{ABC}^2+\sigma_{e_t}^2,\\
	E(MS_{AC})&=br\sigma_{AC}^2+r\sigma_{ABC}^2+\sigma_{e_t}^2,\\
	E(MS_{BC})&=\frac{ar}{(b-1)(c-1)}\sum_{j=1}^{b}\sum_{k=1}^{c}(BC_{jk}-\overline{BC}_{.k}-				\overline{BC}_{j.}+\overline{BC}_{..})^2+r\sigma_{ABC}^2+\sigma_{e_t}^2,\\
	E(MS_{ABC})&=r\sigma_{ABC}^2+\sigma_{e_t}^2.
\end{align*}

\subsection{Expected mean squares when only $B$ is random}\label{Brandom}

In this case, $B$, $AB$, $BC$ and $ABC$ are random with variance $\sigma_B^2$, $\sigma_{AB}^2$, $\sigma_{BC}^2$ and $\sigma_{ABC}^2$, respectively. Thus, we get:

\begin{align*}
	E(MS_A)&=\frac{bcr}{a-1}\sum_{i=1}^{a}(A_i-\overline{A}_{.}+\overline{AC}_{i.}-\overline{AC}_{..})^2
		+bc\sigma_{e_A}^2+cr\sigma_{AB}^2+c\sigma_{e_{AB}}^2+r\sigma_{ABC}^2+\sigma_{e_t}^2,\\
	E(MS_B)&=acr\sigma_{B}^2+ac\sigma_{e_B}^2+cr\sigma_{AB}^2+c\sigma_{e_{AB}}^2
		+ar\sigma_{BC}^2+r\sigma_{ABC}^2+\sigma_{e_t}^2,\\
	E(MS_{AB})&=cr\sigma_{AB}^2+c\sigma_{e_{AB}}^2+r\sigma_{ABC}^2+\sigma_{e_t}^2,\\
	E(MS_C)&=\frac{abr}{c-1}\sum_{k=1}^{c}(C_k-\overline{C}_{.}+\overline{AC}_{.k}-						\overline{AC}_{..})^2+ar\sigma_{BC}^2+r\sigma_{ABC}^2+\sigma_{e_t}^2,\\
	E(MS_{AC})&=\frac{br}{(a-1)(c-1)}\sum_{i=1}^{a}\sum_{k=1}^{c}(AC_{ik}-\overline{AC}_{.k}-					\overline{AC}_{i.}+\overline{AC}_{..})^2+r\sigma_{ABC}^2+\sigma_{e_t}^2,\\
	E(MS_{BC})&=ar\sigma_{BC}^2+r\sigma_{ABC}^2+\sigma_{e_t}^2,\\
	E(MS_{ABC})&=r\sigma_{ABC}^2+\sigma_{e_t}^2.
\end{align*}

\subsection{Expected mean squares when only $C$ is random}\label{Crandom}

Here $C$, $AC$, $BC$ and $ABC$ are random with variance $\sigma_C^2$, $\sigma_{AC}^2$, $\sigma_{BC}^2$ and $\sigma_{ABC}^2$, respectively. Thus, we get:

\begin{align*}
	E(MS_A)&=\frac{bcr}{a-1}\sum_{i=1}^{a}(A_i-\overline{A}_{.}+\overline{AB}_{i.}-\overline{AB}_{..})^2
		+bc\sigma_{e_A}^2+c\sigma_{e_{AB}}^{2}+br\sigma_{AC}^2+r\sigma_{ABC}^2+\sigma_{e_t}^2,\\
	E(MS_B)&=\frac{acr}{b-1}\sum_{j=1}^{b}(B_j-\overline{B}_{.}+\overline{AB}_{.j}-\overline{AB}_{..})^2
		+ac\sigma_{e_B}^2+c\sigma_{e_{AB}}^{2}+ar\sigma_{BC}^2+r\sigma_{ABC}^2
		+\sigma_{e_t}^2,\\
	E(MS_{AB})&=\frac{cr}{(a-1)(b-1)}\sum_{i=1}^{a}\sum_{j=1}^{b}(AB_{ij}-\overline{AB}_{.j}-				\overline{AB}_{i.}+\overline{AB}_{..})^2+r\sigma_{ABC}^2+c\sigma_{e_{AB}}^2+\sigma_{e_t}^2,\\
	E(MS_C)&=abr\sigma_{C}^2+br\sigma_{AC}^2+ar\sigma_{BC}^2+r\sigma_{ABC}^2+\sigma_{e_t}^2
		,\\
	E(MS_{AC})&=br\sigma_{AC}^2+r\sigma_{ABC}^2+\sigma_{e_t}^2,\\
	E(MS_{BC})&=ar\sigma_{BC}^2+r\sigma_{ABC}^2+\sigma_{e_t}^2,\\
	E(MS_{ABC})&=r\sigma_{ABC}^2+\sigma_{e_t}^2.
\end{align*}

\section{$F$ tests}

In this section we present the $F$ tests in tables. When required, we will specify the approximated $df$ by means of the famous estimator developed by \cite{Satter}; for the cases in which the complex estimation is a function of two variance components we will use the estimator proposed by \cite{Ames}, which is a correction to Satterthwaite for this particular case. Let us start with the estimator by Satterthwaite:

If $\theta$ is variance which is a linear combination of $m$ independent variances, i.e., if  $\theta =\sum_{i=1}^ma_i\theta_i$, with estimator $\hat\theta=\sum_{i=1}^ma_iMS_i^2$, we say that $\hat\theta$ is a complex estimator of $\theta$. Since for our case the coefficients $a_i=1$, for $i=1,\ldots, m$, we will omit them on what follows. For the cases in which the variance estimator is complex, \cite{Satter} proposed the following estimator for the $df$:
\begin{align}\label{Satter}
	\hat f_s=\frac{\left(\sum_{i=1}^mMS_i\right)^2}{\sum_{i=1}^mMS_i^2/n_i},
\end{align}
where $n_i$ are the $df$ of the source of variation corresponding to $i$. This is so because $\frac{f\hat\theta}{\theta}$ can be approximated to a $\chi^2$ with $f$ degrees of freedom. This estimator is widely known, so we proceed to present the estimator proposed by \cite{Ames}:

When the variance $\theta$ is a function of two variance components $\theta_1$ and $\theta_2$, call $\phi_1=1$ and $\phi_2=\theta_2/\theta_1$, and consider the class of estimators given by $\hat\phi_2=rMS_2/MS_1$, where $r$ is a constant, then we can approximate the $df$ by
\begin{align}\label{aw}
	\hat f_{aw}(r)=\frac{\left(\sum_{i=1}^2\hat\phi_i\right)^2}{\sum_{i=1}^2\phi_i^2/n_i}.
\end{align}
Note that $\hat f_{aw}(1)=\hat f_s$ and that $\min(n_1,n_2)\leq\hat f_{aw}(r)\leq n_1+n_2$. Thus, we can vary $r$ in order to get  better properties. For instance,
\begin{align*}
	r^*=\frac{n_2}{n_2-2}\left(\frac{2(n_1+n_2-2)}{n_1(n_2-4)}+1\right)
\end{align*}
minimizes the mean square of the error of $1/\hat\phi_2$. Also $r^*>1$ and $\hat f_{aw}(r^*)<f_s$.
In this paper, every time we calculate the Ames-Webster estimator (\ref{aw}), we will also calculate its respective value $r^*$.  Using the Ames-Webster approach we have two possible estimations for every value of $r$. Then, if both of them are less than $\hat f_s$, it is advisable to use the larger one, since the smaller one usually has a negative bias.

With these tools at hand, we proceed to present the $F$ test for every model. The first column in each of the following tables will be the source of variation, the second one will tell us if the effects are random or fixed, the third one will be the corresponding $F$ test and the last one will be the null hypothesis under consideration. When the effects are random, the null hypothesis will be that the corresponding variance of the source has 0 variance; when the effects are fixed, the null hypothesis will be that all effects are equal (to 0).

\subsection{$F$ tests when all effects are fixed}\label{FFixed}

To construct the $F$ tests in Table \ref{tab1}, we use the expected mean squares found in Subsection \ref{fixed}. Note that $R$ will have the same structure for the $F$ test, regardless of it being constant or random (although, of course, the hypothesis will change).
\begin{table}[htpb]
	\centering\small\caption{$F$ tests for the fixed effects model}\label{tab1}
	\begin{tabular}{|c|c|c|c|}\hline
		Source  & Effect &   $F$ & $H_0$\\ \hline
		$R$          &   $f$  & $\frac{MS_R+MS_{e_{AB}}}{MS_{e_A}+MS_{e_B}}$ & $\sigma_R^2=0$\\ \hline	   
		$A$           &   $f$  & $\frac{MS_A}{MS_{e_A}}$                    &  $A_1=A_2=\cdots=A_a=0$\\ \hline
		$e_A$      &   $r$  & $\frac{MS_{e_A}}{MS_{e_{AB}}}$                    & $\sigma_{e_A}^2=0$\\ \hline
		$B$           &   $f$  & $\frac{MS_B}{MS_{e_B}}$                    & $B_1=B_2=\cdots=B_b=0$\\ \hline
		$e_B$      &   $r$  & $\frac{MS_{e_B}}{MS_{e_{AB}}}$                    & $\sigma_{e_B}^2=0$\\ \hline
		$AB$        &   $f$  & $\frac{MS_{AB}}{MS_{e_{AB}}}$                    & $(AB)_{ij}=0$, $\forall i$, $\forall j$.\\ \hline
		$e_{AB}$ &   $r$  & $\frac{MS_{e_{AB}}}{MS_{e_t}}$              & $\sigma_{e_{AB}}^2=0$\\ \hline
		$C$           &   $f$  & $\frac{MS_C}{MS_{e_t}}$              & $C_1=C_2=\cdots=C_c=0$\\ \hline
		$AC$        &   $f$  & $\frac{MS_{AC}}{MS_{e_t}}$              & $(AC)_{ik}=0$, $\forall i$, $\forall k$.\\ \hline
		$BC$        &   $f$  & $\frac{MS_{BC}}{MS_{e_t}}$         & $(BC)_{jk}=0$, $\forall j$, $\forall k$.\\ \hline
		$ABC$     &   $f$  & $\frac{MS_{ABC}}{MS_{e_t}}$         & $(ABC)_{ijk}=0$, $\forall i$, $\forall j$, $\forall 			k$.\\ \hline
		$e_t$       &   $r$  & -- &                  \\ \hline
	\end{tabular}	
\end{table}

Using the Satterthwaite estimator in (\ref{Satter}), we approximate the $df$ for $R$ as:
\begin{align*}
	v_1&=\frac{(MS_R+MS_{e_{AB}})^2}{\frac{MS_R^2}{r-1}+\frac{MS_{e_{AB}}^2}{(r-1)(a-1)(b-1)}},\\
	v_2&=\frac{(MS_{e_A}+MS_{e_B})^2}{\frac{MS_{e_A}^2}{(r-1)(a-1)}+\frac{MS_{e_B}^2}{(r-1)(b-1)}},
\end{align*}
where $v_1$ and $v_2$ are the $df$ for the enumerator and the denominator, respectively.

When we adjust using the Ames-Webster estimator (\ref{aw}), we obtain two estimators for each case. First let us see the $df$ for the numerator: Let $MS_1=MS_R$ and $MS_2=MS_{e_{AB}}$, then

\begin{align*}
	p_1=\frac{(r-1)(a-1)(b-1)}{(r-1)(a-1)(b-1)-2}\left(\frac{2[(r-1)(a-1)(b-1)+r-3]}{(r-1)[(r-1)(a-1)(b-1)-4]}			+1\right),
\end{align*}
\begin{align*}
	\hat f_{aw}(p_1)=\frac{(1+p_1MS_{e_{AB}}/MS_R)^2}{\frac{1}{r-1}+\frac{(p_1MS_{e_{AB}}/MS_R)^2}{(r-1)(a-1)(b-1)}};
\end{align*}

\noindent on the other hand, when $MS_1=MS_{e_{AB}}$ and $MS_2=MS_R$:
\begin{align*}
	p_1^*&= \frac{r-1}{r-3}\left(\frac{2[(r-1)(a-1)(b-1)+r-3]}{(r-1)(a-1)(b-1)(r-5)}+1\right),
\end{align*}
\begin{align*}
	\hat f_{aw}(p_1^*)&=\frac{(1+p_1^*MS_R/MS_{e_{AB}})^2}{\frac{1}{(r-1)(a-1)(b-1)}+\frac{(p_1^*MS_R/MS_{e_{AB}})^2}			{r-1}}.
\end{align*}
Now, for the denominator, when $MS_1=MS_{e_A}$ and $MS_2=MS_{e_B}$ we have:
\begin{align*}
	p_2=\frac{(r-1)(b-1)}{(r-1)(b-1)-2}\left(\frac{2\{(r-1)[(a-1)+(b-1)]-2\}}{(r-1)(a-1)[(r-1)(b-1)-4]}+1\right),
\end{align*}
\begin{align*}
	\hat f_{aw}(p_2)=\frac{(1+p_2MS_{e_B}/MS_{e_A})^2}{\frac{1}{(r-1)(a-1)}+\frac{(p_2MS_{e_B}/M-3)^2}{(r-1)(b-1)}};
\end{align*}

\noindent and when $MS_1=MS_{e_B}$ and $MS_2=MS_{e_A}$ we have:
\begin{align*}
	p_2^*=\frac{(r-1)(a-1)}{(r-1)(a-1)-2}\left(\frac{2\{(r-1)[(a-1)+(b-1)]-2\}}{(r-1)(b-1)[(r-1)(a-1)-4]}+1\right),
\end{align*}
\begin{align*}
	\hat f_{aw}(p_2^*)=\frac{(1+p_2^*MS_{e_A}/MS_{e_B})^2}{\frac{1}{(r-1)(b-1)}+\frac{(p_2^*MS_{e_A}/M-5)^2}{(r-1)			(a-1)}}.
\end{align*}

The estimators for the $df$ of $R$ will always be the same. For this reason they will be omitted on the tables to come.

\subsection{$F$ tests when all effects are random}\label{Frandom}

\noindent When all effects are random, we construct the $F$ tests on Table \ref{tab2} based on the mean squares developed in Subsection \ref{random}.

\begin{table}[htpb]
	\centering\small\caption{$F$ tests for the random effects model}\label{tab2}
	\begin{tabular}{|c|c|c|c|}\hline
		Source       & Effect &   $F$ & $H_0$\\ \hline
		$R$            &   $r$    & $\frac{MS_R+MS_{e_{AB}}}{MS_{e_A}+MS_{e_B}}$                          & $\sigma_R^2=0$\\ \hline	   
		$A$            &   $r$    & $\frac{MS_A+MS_{e_{AB}}+MS_{ABC}}{MS_{e_A}+MS_{AB}+MS_{AC}}$ & $\sigma_A^2=0$\\ \hline
		$e_A$       &   $r$    & $\frac{MS_{e_A}}{MS_{e_{AB}}}$                                                & $\sigma_{AR}^2=0$\\ \hline
		$B$            &   $r$    & $\frac{MS_B+MS_{e_{AB}}+MS_{ABC}}{MS_{e_B}+MS_{AB}+MS_{BC}}$ & $\sigma_B^2=0$\\ \hline
		$e_B$       &   $r$    & $\frac{MS_{e_B}}{MS_{e_{AB}}}$                                                & $\sigma_{BR}^2=0$\\ \hline
		$AB$         &   $r$    & $\frac{MS_{AB}+MS_{e_t}}{MS_{e_{AB}}+MS_{ABC}}$                 & $\sigma_{AB}^2=0$\\ \hline
		$e_{AB}$ &   $r$     & $\frac{MS_{e_{AB}}}{MS_{e_t}}$                                    & $\sigma_{e_{AB}}^2=0$\\ \hline
		$C$           & $r$       & $\frac{MS_C+MS_{ABC}}{MS_{AC}+MS_{BC}}$                & $\sigma_C^2=0$\\ \hline
		$AC$        & $r$       & $\frac{MS_{AC}}{MS_{ABC}}$                                         & $\sigma_{AC}^2=0$\\ \hline
		$BC$        & $r$       & $\frac{MS_{BC}}{MS_{ABC}}$                                    & $\sigma_{BC}^2=0$\\ \hline
		$ABC$     & $r$       & $\frac{MS_{ABC}}{MS_{e_t}}$                                   & $\sigma_{ABC}^2=0$\\ \hline
		$e_t$        &   $r$     & --                                                                            &                \\ \hline
\end{tabular}
\end{table}

Since the complex estimators for effects $A$ and $B$ in Table \ref{tab2} have three variance components, we will use only (\ref{Satter}) with them to find their approximate $df$.  For the effects of $A$, the $df$ in the numerator and the denominator $v_1$ and $v_2$, respectively, will be given by:
\begin{align}\label{SatArandom}
	v_1&=\frac{(MS_A+MS_{e_{AB}}+MS_{ABC})^2}{\frac{MS_A^2}{a-1}+\frac{MS_{e_{AB}}^2}{(a-1)(b-1)(r-1)}+\frac{MS_{ABC}^2}		{(a-1)(b-1)(c-1)}},\notag \\
	v_2&=\frac{(MS_{e_A}+MS_{AB}+MS_{AC})^2}{\frac{MS_{e_A}^2}{(r-1)(a-1)}+\frac{MS_{AB}^2}{(a-1)(b-1)}+\frac{MS_{AC}^2}{(a-1)		(c-1)}}.
\end{align}

With $B$, the $df$ of the $F$ test will be for the numerator and denominator  respectively:
\begin{align}\label{SatBrandom}
	v_1&=\frac{(MS_B+MS_{e_{AB}}+MS_{ABC})^2}{\frac{MS_B^2}{b-1}+\frac{MS_{e_{AB}}^2}{(a-1)(b-1)(r-1)}+\frac{MS_{ABC}^2}		{(a-1)(b-1)(c-1)}},\notag \\
	v_2&=\frac{(MS_{e_B}+MS_{AB}+MS_{BC})^2}{\frac{MS_{e_B}^2}{(r-1)(b-1)}+\frac{MS_{AB}^2}{(a-1)(b-1)}+\frac{MS_{BC}^2}		{(b-1)(c-1)}}.
\end{align}

The $df$ for $AB$ approximated by (\ref{Satter}) will be respectively for the numerator and the denominator:
\begin{align}\label{SatterABrandom}
	v_1&=\frac{(MS_{AB}+MS_{e_t})^2}{\frac{MS_{AB}^2}{(a-1)(b-1)}+\frac{MS_{e_t}^2}{ab(c-1)(r-1)}},\notag \\
	v_2&=\frac{(MS_{e_{AB}}+MS_{ABC})^2}{\frac{MS_{e_{AB}}^2}{(a-1)(b-1)(r-1)}+\frac{MS_{ABC}^2}{(a-1)(b-1)(c-1)}}.
\end{align}
Adjusting by means of (\ref{aw}), there are two possible estimators in each case. First, let us see the degrees of freedom in the numerator. Let $MS_1=MS_{AB}$ and $MS_2=MS_{e_t}$, then
\begin{align*}
	p_1=\frac{ab(c-1)(r-1)}{ab(c-1)(r-1)-2}&\left(\frac{2[(a-1)(b-1)+ab(c-1)(r-1)-2]}{(a-1)(b-1)[ab(c-1)			(r-1)-4]}+1\right),
\end{align*}
\begin{align}\label{AWABrandom}
	&\hat f_{aw}(p_1)=\frac{(1+p_1MS_{e_t}/MS_{AB})^2}{\frac{1}{(a-1)(b-1)}+\frac{(p_1MS_{e_t}/MS_{AB})^2}{ab(c-1)		(r-1)}};
\end{align}
on the other hand, when $MS_1=MS_{e_t}$ and $MS_2=MS_{AB}$:
\begin{align*}
	p_1^*= \frac{(a-1)(b-1)}{(a-1)(b-1)-2}\left(\frac{2[(a-1)(b-1)+ab(c-1)(r-1)-2]}{ab(c-1)(r-1)[(a-1)				(b-1)-4]}+1\right),
\end{align*}
\begin{align}\label{AWABrrandom}
	\hat f_{aw}(p_1^*)=\frac{(1+p_1^*MS_{AB}/MS_{e_t})^2}{\frac{1}{ab(c-1)(r-1)}+\frac{(p_1^*MS_{AB}/				MS_{e_t})^2}{(a-1)(b-1)}}.
\end{align}
And for the denominator, when $MS_1=MS_{e_{AB}}$ and $MS_2=MS_{ABC}$ we have:
\begin{align*}
	p_2=\frac{(a-1)(b-1)(c-1)}{(a-1)(b-1)(c-1)-2}\left(\frac{2[(a-1)(b-1)(c+r-2)-2]}{(a-1)(b-1)(r-1)[(a-1)(b-1)		(c-1)-4]}+1\right),
\end{align*}
\begin{align}\label{AWABrandomm}
	\hat f_{aw}(p_2)=\frac{(1+p_2MS_{ABC}/MS_{e_{AB}})^2}{\frac{1}{(a-1)(b-1)(r-1)}+\frac{(p_2MS_{ABC}/MS_{e_{AB}})^2}{(a-1)		(b-1)(c-1)}};
\end{align}
when $MS_1=MS_{ABC}$ and $MS_2=MS_{e_{AB}}$ we get:
\begin{align*}
	p_2^*=\frac{(a-1)(b-1)(r-1)}{(a-1)(b-1)(r-1)-2}\left(\frac{2[(a-1)(b-1)(c+r-2)-2]}{(a-1)(b-1)(c-1)[(a-1)		(b-1)(r-1)-4]}+1\right),
\end{align*}
\begin{align}\label{AWABrrandomm}
	\hat f_{aw}(p_2^*)=\frac{(1+p_2^*MS_{e_{AB}}/MS_{ABC})^2}{\frac{1}{(a-1)(b-1)(c-1)}+							\frac{(p_2^*MS_{e_{AB}}/MS_{ABC})^2}{(a-1)(b-1)(r-1)}}.
\end{align}

With $C$, using $\hat f_s$, the $df$ for the numerator and denominator will be respectively:
\begin{align}\label{SatCrandom}
	v_1&=\frac{(MS_C+MS_{ABC})^2}{\frac{MS_C^2}{c-1}+\frac{MS_{ABC}^2}{(a-1)(b-1)(c-1)}},\notag\\
	v_2&=\frac{(MS_{AC}+MS_{BC})^2}{\frac{MS_{AC}^2}{(a-1)(c-1)}+\frac{MS_{ABC}^2}{(a-1)(b-1)(c-1)}}.
\end{align}

For $\hat f_{aw}$ these were the estimators for the $df$ of the numerator when $MS_1=MS_C$ and $MS_2=MS_{ABC}$:
\begin{align*}
	p_1=\frac{(a-1)(b-1)(c-1)}{(a-1)(b-1)(c-1)-2}\left(\frac{2[(a-1)(b-1)(c-1)+c-3]}{(c-1)[(a-1)(b-1)(c-1)-4]}			+1\right),
\end{align*}
\begin{align}\label{AWCrandom}
	\hat f(p_1)=\frac{(1+p_1MS_{ABC}/MS_C)^2}{\frac{1}{c-1}+\frac{(p_1MS_{ABC}/MS_C)^2}{(a-1)(b-1)(c-1)}};
\end{align}
still for the numerator, but exchanging the order of $MS_1$ and $MS_2$, we obtain:
\begin{align*}
	p_1^*=\frac{c-1}{c-3}\left(\frac{2[(a-1)(b-1)(c-1)+c-3]}{(a-1)(b-1)(c-1)(c-5)}+1\right),
\end{align*}
\begin{align}\label{AWCrrandom}
	\hat f(p_1^*)=\frac{(1+p_1^*MS_C/MS_{ABC})^2}{\frac{1}{(a-1)(b-1)(c-1)}+\frac{(p_1^*MS_C/					MS_{ABC})^2}{c-1}}.
\end{align}
For the denominator, taking $MS_1=MS_{AC}$ and $MS_2=MS_{BC}$, we get the following estimations:
\begin{align*}
	p_2=\frac{(b-1)(c-1)}{(b-1)(c-1)-2}\left(\frac{2\{(c-1)[(a-1)+(b-1)]-2\}}{(a-1)(c-1)[(b-1)(c-1)-4]}+1\right)
\end{align*}
\begin{align}\label{AWCrandomm}
	\hat f(p_2)=\frac{(1+p_2MS_{BC}/MS_{AC})^2}{\frac{1}{(a-1)(c-1)}+\frac{(p_2MS_{BC}/MS_{AC})^2}{(b-1)(c-1)}};
\end{align}
once again for the denominator, but exchanging to $MS_1=MS_{BC}$ and $MS_2=MS_{AC}$, we get:
\begin{align*}
	p_2^*=\frac{(a-1)(c-1)}{(a-1)(c-1)-2}\left(\frac{2\{(c-1)[(a-1)+(b-1)]-2\}}{(b-1)(c-1)[(a-1)(c-1)-4]}+1\right),
\end{align*}
\begin{align}\label{AWCrrandomm}
	\hat f(p_2^*)=\frac{(1+p_2^*MS_{AC}/MS_{BC})^2}{\frac{1}{(b-1)(c-1)}+\frac{(p_2^*MS_{AC}/MS_{BC})^2}{(a-1)			(c-1)}}.
\end{align}

\subsection{$F$ tests when only one factor has fixed effects}

With respect to Table \ref{tab2}, the only difference for the three cases considered here (only $A$ has fixed effects, only $B$ has fixed effects, and only $C$ has fixed effects) will occur in the row corresponding to the fixed effect: first, obviously, its effect will be $f$ instead of $r$; second, its null hypothesis will be about the equality of all treatments inside that factor. So when $A$ is the only factor of fixed effects, its effect is $f$ and its null hypothesis is $A_1=\cdots=A_a=0$, all other fields remaining equal to Table \ref{tab2}; when $B$ is the only factor with fixed effects, its effect is $f$ and its null hypothesis is $B_1=\cdots=B_b=0$,  all other fields remaining equal to Table \ref{tab2}; and when the only fixed effects are those corresponding to $C$, its value at effect is $f$ and the null hypothesis will be $C_1=\cdots=C_c=0$,  all other fields remaining equal to Table \ref{tab2}. This can be easily verified with the information in Subsections \ref{Afixed}, \ref{Bfixed} and \ref{Cfixed}

Since, in particular, the structure of the complex variance estimators is identical to the structure of the model with random effects, the approximate $df$ for each of these three cases are exactly the same to those found in Subsection \ref{Frandom}.

\subsection{$F$ tests when only $A$ has random effects}

When the effects of $A$ are random, we obtain Table \ref{tab3} based on the $E(MS)$'s found in Subsection \ref{Arandom}. 

\begin{table}[htpb]
	\centering\small\caption{$F$ tests when only $A$ is random}\label{tab3}
	\begin{tabular}{|c|c|c|c|}\hline
	Fuente     & Efecto &   $F$ & $H_0$\\ \hline
$R$        &   $r$  & $\frac{MS_R+MS_{e_{AB}}}{MS_{e_A}+MS_{e_B}}$ & $\sigma_R^2=0$\\ \hline	   
$A$        &   $r$  & $\frac{MS_A+MS_{e_{AB}}+MS_{ABC}}{MS_{e_A}+MS_{AB}+MS_{AC}}$ & $\sigma_A^2=0$\\ \hline
$e_A$      &   $r$  & $\frac{MS_{e_A}}{MS_{e_{AB}}}$ & $\sigma_{e_A}^2=0$\\ \hline
$B$        &   $f$  & $\frac{MS_B+MS_{e_{AB}}}{MS_{e_B}+MS_{AB}}$ & $B_1=B_2=\cdots=B_b=0$\\ \hline
$e_B$      &   $r$  & $\frac{MS_{e_B}}{MS_{e_{AB}}}$ & $\sigma_{e_B}^2=0$\\ \hline
$AB$       &   $r$  & $\frac{MS_{AB}+MS_{e_t}}{MS_{e_{AB}}+MS_{ABC}}$ & $\sigma_{AB}^2=0$\\ \hline
$e_{AB}$   &   $r$  & $\frac{MS_{e_{AB}}}{MS_{e_t}}$ & $\sigma_{e_{AB}}^2=0$\\ \hline
$C$        &   $f$  & $\frac{MS_C}{MS_{AC}}$ & $C_1=C_2=\cdots=C_c=0$\\ \hline
$AC$       &   $r$  & $\frac{MS_{AC}}{MS_{ABC}}$ & $\sigma_{AC}^2=0$\\ \hline
$BC$       &   $f$  & $\frac{MS_{BC}}{MS_{ABC}}$ & $(BC)_{jk}=0$, $\forall j,k$.\\ \hline
$ABC$      &   $r$  & $\frac{MS_{ABC}}{MS_{e_t}}$ & $\sigma_{ABC}^2=0$\\ \hline
$e_t$ &   $r$  & --    &               \\ \hline
\end{tabular}	
\end{table}

The estimators for the $df$ of $A$ are those in (\ref{SatArandom}). The estimators of the $df$ by Satterthwaite for $AB$ are those in (\ref{SatterABrandom}); the estimators by Ames-Webster are given in equations (\ref{AWABrandom}) and (\ref{AWABrrandom}) for the numerator, and (\ref{AWABrandomm}) and (\ref{AWABrrandomm}) for the denominator. Now we procede to evaluate the $df$ for the $F$ test of $B$, first by means of the Satterthwaite estimator in equation (\ref{Satter}):

\begin{align}
	v_1&=\frac{(MS_B+MS_{e_{AB}})^2}{\frac{MS_B^2}{b-1}+\frac{MS_{e_{AB}}^2}{(a-1)(b-1)(r-1)}},\notag \\
	v_2&=\frac{(MS_{e_B}+MS_{AB})^2}{\frac{MS_{e_B}^2}{(b-1)(r-1)}+\frac{MS_{AB}^2}{(a-1)(b-1)}}.
\end{align}
Still with $B$, the first Ames-Webster estimator for the $df$ of the numerator of the $F$, taking $MS_1=MS_B$ and $MS_2=MS_{e_{AB}}$ will be:
\begin{align*}
	p_1&=\frac{(a-1)(b-1)(r-1)}{(a-1)(b-1)(r-1)-2} \left(\frac{2\{(b-1)[(a-1)(r-1)+1]-2\}}{(b-1)[(a-1)(b-1)(r-1)-4]}		+1\right),
\end{align*}
\begin{align*}
	\hat f(p_1)&=\frac{(1+p_1MS_{e_{AB}}/MS_B)^2}{\frac{1}{b-1}+\frac{(p_1MS_{e_{AB}}/MS_B)^2}{(a-1)(b-1)(r-1)}};
\end{align*}
and exchanging the order to $MS_1=MS_{e_{AB}}$ and $MS_2=MS_B$, we obtain:
\begin{align*}
	p_1^*=\frac{b-1}{b-3}\left(\frac{2\{(b-1)[(a-1)(r-1)+1]-2\}}{(a-1)(b-1)(r-1)(b-5)}+1\right),
\end{align*}
\begin{align*}
	\hat f(p_1^*)=\frac{(1+p_1^*MS_B/MS_{e_{AB}})^2}{\frac{1}{(a-1)(b-1)(r-1)}+\frac{(p_1^*MS_B/MS_{e_{AB}})^2}{b-1}}.
\end{align*}
For the denominator of the the $F$ test of $B$, taking $MS_1=MS_{e_B}$ y $MS_2=MS_{AB}$:
\begin{align*}
	p_2&=\frac{(a-1)(b-1)}{(a-1)(b-1)-2}\left(\frac{2[(b-1)(a+r-2)-2]}{(b-1)(r-1)[(a-1)(b-1)-4]}+1\right),
\end{align*}
\begin{align*}
	\hat f(p_2)&=\frac{(1+p_2MS_{AB}/MS_{e_B})^2}{\frac{1}{(b-1)(r-1)}+\frac{(p_2MS_{AB}/MS_{e_B})^2}{(a-1)(b-1)}};
\end{align*}
and exchanging the order of $MS_1$ and $MS_2$:
\begin{align*}
	p_2^*=\frac{(b-1)(r-1)}{(b-1)(r-1)-2}\left(\frac{2[(b-1)(a+r-2)-2]}{(a-1)(b-1)[(b-1)(r-1)-4]}+1\right),
\end{align*}
\begin{align*}
	\hat f(p_2^*)=\frac{(1+p_2^*MS_{e_B}/MS_{AB})^2}{\frac{1}{(a-1)(b-1)}+\frac{(p_2^*MS_{e_B}/MS_{AB})^2}{(b-1)(r-1)}}.
\end{align*}

\subsection{$F$ tests when only $B$ has random effects}

When only $A$ and $C$ have fixed effects, based on Subsection (\ref{Brandom}), we get the following:

\begin{table}[htpb]
	\centering\small\caption{$F$ tests when only $B$ is random}\label{tab4}
	\begin{tabular}{|c|c|c|c|}\hline
		Source      & Effect  &   $F$ & $H_0$\\ \hline
		$R$           &   $r$  & $\frac{MS_R+MS_{e_{AB}}}{MS_{e_A}+MS_{e_B}}$ & $\sigma_R^2=0$\\ \hline	   
		$A$           &   $f$  & $\frac{MS_A+MS_{e_{AB}}}{MS_{e_A}+MS_{AB}}$ & $A_1=A_2=\cdots=A_a=0$\\ \hline
		$e_A$      &   $r$  & $\frac{MS_{e_A}}{MS_{e_{AB}}}$ & $\sigma_{e_A}^2=0$\\ \hline
		$B$           &   $r$  & $\frac{MS_B+MS_{e_{AB}}+MS_{ABC}}{MS_{e_B}+MS_{AB}+MS_{BC}}$ & $\sigma_B^2=0$\\ \hline
		$e_B$      &   $r$  & $\frac{MS_{e_B}}{MS_{e_{AB}}}$ & $\sigma_{e_B}^2=0$\\ \hline
		$AB$        &   $r$  & $\frac{MS_{AB}+MS_{e_t}}{MS_{e_{AB}}+MS_{ABC}}$ & $\sigma_{AB}^2=0$\\ \hline
		$e_{AB}$ &   $r$  & $\frac{MS_{e_{AB}}}{MS_{e_t}}$ & $\sigma_{e_{AB}}^2=0$\\ \hline
		$C$          &   $f$   & $\frac{MS_C}{MS_{BC}}$ & $C_1=C_2=\cdots=C_c=0$\\ \hline
		$AC$        &   $f$  & $\frac{MS_{AC}}{MS_{ABC}}$ & $(AC)_{ik}=0$, $\forall i,k$.\\ \hline
		$BC$        &   $r$  & $\frac{MS_{BC}}{MS_{ABC}}$ & $\sigma_{BC}^2=0$\\ \hline
		$ABC$      &   $r$   & $\frac{MS_{ABC}}{MS_{e_t}}$ & $\sigma_{ABC}^2=0$\\ \hline
		$e_t$ &   $r$  & --  &                 \\ \hline
	\end{tabular}	
\end{table}

The approximated $df$ for the $F$ test of $B$ were found using the Satterthwaite estimator (\ref{SatBrandom}). The approximation of the $df$ for $AB$ using Satterthwaite is given by (\ref{SatterABrandom}); using Ames-Webster, the estimator for the $df$ of $AB$ are given in equations (\ref{AWABrandom}) and (\ref{AWABrrandom}) for the numerator, and (\ref{AWABrandomm}) and (\ref{AWABrrandomm}) for the denominator. Now we procede to evaluate the $df$ for the $F$ test of $A$, first by means of the Satterthwaite estimator in equation (\ref{Satter}):

\begin{align*}
	v_1&=\frac{(MS_A+MS_{e_{AB}})^2}{\frac{MS_A^2}{a-1}+\frac{MS_{e_{AB}}^2}{(a-1)(b-1)(r-1)}},\\
	v_2&=\frac{(MS_{e_A}+MS_{AB})^2}{\frac{MS_{e_A}^2}{(r-1)(a-1)}+\frac{MS_{AB}^2}{(a-1)(b-1)}}.
\end{align*}
The Ames-Webster estimator for the numerator is the following when $MS_1=MS_A$ and $MS_2=MS_{e_{AB}}$:
\begin{align*}
	p_1=\frac{(a-1)(b-1)(r-1)}{(a-1)(b-1)(r-1)-2}\left(\frac{2\{(a-1)[(b-1)(r-1)+1]-2\}}{(a-1)[(a-1)(b-1)(r-1)-4]}		+1\right),
\end{align*}
\begin{align*}
	\hat f(p_1)=\frac{(1+p_1MS_{e_{AB}}/MS_A)^2}{\frac{1}{a-1}+\frac{(p_1MS_{e_{AB}}/MS_A)^2}{(a-1)(b-1)(r-1)}};
\end{align*}
still with the numerator but taking $MS_1=MS_{e_{AB}}$ and $MS_2=MS_A$, we get:
\begin{align*}
	p_1^*=\frac{a-1}{a-3}\left(\frac{2\{(a-1)[(b-1)(r-1)+1]-2\}}{(a-1)(b-1)(r-1)(a-5)}+1\right),
\end{align*}
\begin{align*}
	\hat f(p_1^*)=\frac{(1+p_1^*MS_A/MS_{e_{AB}})^2}{\frac{1}{(a-1)(b-1)(r-1)}+\frac{(p_1^*MS_A/MS_{e_{AB}})^2}{a-1}}.
\end{align*}

\noindent For the denominator, doing $MS_1=MS_{e_A}$ and $MS_2=MS_{AB}$, we get:
\begin{align*}
	p_2=\frac{(a-1)(b-1)}{(a-1)(b-1)-2}\left(\frac{2[(a-1)(b+r-2)-2]}{(a-1)(r-1)[(a-1)(b-1)-4]}+1\right),
\end{align*}
\begin{align*}
	\hat f(p_2)&=\frac{(1+p_2MS_{AB}/MS_{e_A})^2}{\frac{1}{(a-1)(r-1)}+\frac{(p_2MS_{AB}/MS_{e_A})^2}{(a-1)(b-1)}};
\end{align*}
finally, exchanging the order of $MS_1$ and $MS_2$, we obtain:
\begin{align*}
	p_2^*=\frac{(a-1)(r-1)}{(a-1)(r-1)-2}\left(\frac{2[(a-1)(b+r-2)-2]}{(a-1)(b-1)[(a-1)(r-1)-4]}+1\right),
\end{align*}
\begin{align*}
	\hat f(p_2^*)=\frac{(1+p_2^*MS_{e_A}/MS_{AB})^2}{\frac{1}{(a-1)(b-1)}+\frac{(p_2^*MS_{e_A}/MS_{AB})^2}{(a-1)(r-1)}}.
\end{align*}

\subsection{$F$ tests when only $C$ has random effects}

Table \ref{tab5} below was constructed using the $E(MS)$'s in Subsection \ref{Crandom}. Note that the  approximate degrees of freedom for $AB$ were described in equations (\ref{SatterABrandom}) by means of Satterthwaite; also for $AB$, the approximations of its degrees of freedom using Ames-Webster were given in equations (\ref{AWABrandom}) and (\ref{AWABrrandom}) for the numerator , and (\ref{AWABrandomm}) and \ref{AWABrrandomm} for the denominator.

For $C$, its approximate $df$ using Satterthwaite were found in (\ref{SatCrandom}). And the Ames-Webster estimators of the $df$ of $C$ are given by (\ref{AWCrandom}) and (\ref{AWCrrandom}) for the numerator, and by (\ref{AWCrandomm}) and (\ref{AWCrrandomm}) for the denominator.

\begin{table}[htpb]
	\centering\small\caption{$F$ tests when only $C$ is random}\label{tab5}
	\begin{tabular}{|c|c|c|c|}\hline
		Source      & Effect &   $F$                                                       & $H_0$\\ \hline
		$R$           &   $r$  & $\frac{MS_R+MS_{e_{AB}}}{MS_{e_A}+MS_{e_B}}$           & $\sigma_R^2=0$\\ \hline	   
		$A$            &   $f$  & $\frac{MS_A+MS_{e_t}}{MS_{e_A}+MS_{AC}}$      & $A_1=A_2=\cdots=A_a=0$\\ \hline
		$e_A$        &   $r$  & $\frac{MS_{e_A}}{MS_{e_{AB}}}$                               & $\sigma_{e_A}^2=0$\\ \hline
		$B$            &   $f$  & $\frac{MS_B+MS_{e_t}}{MS_{e_B}+MS_{BC}}$ & $B_1=B_2=\cdots=B_b=0$\\ \hline
		$e_B$        &   $r$  & $\frac{MS_{e_B}}{MS_{e_{AB}}}$                               & $\sigma_{e_B}^2=0$\\ \hline
		$AB$          &   $f$  & $\frac{MS_{AB}+MS_{e_t}}{MS_{e_{AB}}+MS_{ABC}}$ & $(AB)_{ij}=0$, $\forall i,j$.\\ \hline
		$e_{AB}$   &   $r$  & $\frac{MS_{e_{AB}}}{MS_{e_t}}$                          & $\sigma_{e_{AB}}^2=0$\\ \hline
		$C$            &   $r$  & $\frac{MS_C+MS_{ABC}}{MS_{AC}+MS_{BC}}$ & $\sigma_C^2=0$\\ \hline
		$AC$         &   $r$  & $\frac{MS_{AC}}{MS_{ABC}}$                           & $\sigma_{AC}^2=0$\\ \hline
		$BC$         &   $r$  & $\frac{MS_{BC}}{MS_{ABC}}$                      & $\sigma_{BC}^2=0$\\ \hline
		$ABC$      &   $r$  & $\frac{MS_{ABC}}{MS_{e_t}}$                      & $\sigma_{ABC}^2=0$\\ \hline
		$e_t$        &   $r$  & --                                                                &               \\ \hline
	\end{tabular}	
\end{table}

We procede to evaluate the approximate $df$ for the $F$ test of $A$, first by means of the Satterthwaite estimator in equation (\ref{Satter}):

\begin{align*}
	v_1&=\frac{(MS_A+MS_{e_t})^2}{\frac{MS_A^2}{a-1}+\frac{MS_{e_t}^2}{ab(c-1)(r-1)}},\\
	v_2&=\frac{(MS_{e_A}+MS_{AC})^2}{\frac{MS_{e_A}^2}{(r-1)(a-1)}+\frac{MS_{AC}^2}{(a-1)(c-1)}}.
\end{align*}
The Ames-Webster estimator for the numerator is the following when $MS_1=MS_A$ and $MS_2=MS_{e_t}$:
\begin{align*}
	p_1=\frac{ab(c-1)(r-1)}{ab(c-1)(r-1)-2}\left(\frac{2[ab(c-1)(r-1)+a-3]}{(a-1)[ab(c-1)(r-1)-4]}+1\right),
\end{align*}
\begin{align*}
	\hat f(p_1)=\frac{(1+p_1MS_{e_t}/MS_A)^2}{\frac{1}{a-1}+\frac{(p_1MS_{e_t}/MS_A)^2}{ab(c-1)(r-1)}};
\end{align*}
still with the numerator but taking $MS_1=MS_{e_t}$ and $MS_2=MS_A$, we get:
\begin{align*}
	p_1^*=\frac{a-1}{a-3}\left(\frac{2[ab(c-1)(r-1)+a-3]}{ab(c-1)(r-1)(a-5)}+1\right),
\end{align*}
\begin{align*}
	\hat f(p_1^*)=\frac{(1+p_1^*MS_A/MS_{e_t})^2}{\frac{1}{ab(c-1)(r-1)}+\frac{(p_1^*MS_A/MS_{e_t})^2}{a-1}}.
\end{align*}
For the denominator, doing $MS_1=MS_{e_A}$ and $MS_2=MS_{AC}$, we get:
\begin{align*}
	p_2=\frac{(a-1)(c-1)}{(a-1)(c-1)-2}\left(\frac{2[(a-1)(r+c-2)-2]}{(a-1)(r-1)[(a-1)(c-1)-4]}+1\right),
\end{align*}
\begin{align*}
	\hat f(p_2)&=\frac{(1+p_2MS_{AC}/MS_{e_A})^2}{\frac{1}{(a-1)(r-1)}+\frac{(p_2MS_{AC}/MS_{e_A})^2}{(a-1)(c-1)}};
\end{align*}
finally, exchanging the order of $MS_1$ and $MS_2$, we obtain:
\begin{align*}
	p_2^*=\frac{(a-1)(r-1)}{(a-1)(r-1)-2}\left(\frac{2[(a-1)(r+c-2)-2]}{(a-1)(c-1)[(a-1)(r-1)-4]}+1\right),
\end{align*}
\begin{align*}
	\hat f(p_2^*)=\frac{(1+p_2^*MS_{e_A}/MS_{AC})^2}{\frac{1}{(a-1)(c-1)}+\frac{(p_2^*MS_{e_A}/MS_{AC})^2}{(a-1)(r-1)}}.
\end{align*}

Now, we evaluate the approximate $df$ for the $F$ test of $B$, first by means of the Satterthwaite estimator in equation (\ref{Satter}):

\begin{align*}
	v_1&=\frac{(MS_B+MS_{e_t})^2}{\frac{MS_B^2}{b-1}+\frac{MS_{e_t}^2}{ab(c-1)(r-1)}},\\
	v_2&=\frac{(MS_{e_B}+MS_{BC})^2}{\frac{MS_{e_B}^2}{(r-1)(b-1)}+\frac{MS_{BC}^2}{(b-1)(c-1)}}.
\end{align*}
The Ames-Webster estimator for the numerator is the following when $MS_1=MS_B$ and $MS_2=MS_{e_t}$:
\begin{align*}
	p_1=\frac{ab(c-1)(r-1)}{ab(c-1)(r-1)-2}\left(\frac{2[ab(c-1)(r-1)+b-3]}{(b-1)[ab(c-1)(r-1)-4]}+1\right),
\end{align*}
\begin{align*}
	\hat f(p_1)=\frac{(1+p_1MS_{e_t}/MS_B)^2}{\frac{1}{b-1}+\frac{(p_1MS_{e_t}/MS_B)^2}{ab(c-1)(r-1)}};
\end{align*}
still with the numerator but taking $MS_1=MS_{e_t}$ and $MS_2=MS_B$, we get:
\begin{align*}
	p_1^*=\frac{b-1}{b-3}\left(\frac{2[ab(c-1)(r-1)+b-3]}{ab(c-1)(r-1)(b-5)}+1\right),
\end{align*}
\begin{align*}
	\hat f(p_1^*)=\frac{(1+p_1^*MS_B/MS_{e_t})^2}{\frac{1}{ab(c-1)(r-1)}+\frac{(p_1^*MS_B/MS_{e_t})^2}{b-1}}.
\end{align*}
For the denominator, doing $MS_1=MS_{e_B}$ and $MS_2=MS_{BC}$, we get:
\begin{align*}
	p_2=\frac{(b-1)(c-1)}{(b-1)(c-1)-2}\left(\frac{2[(b-1)(r+c-2)-2]}{(b-1)(r-1)[(b-1)(c-1)-4]}+1\right),
\end{align*}
\begin{align*}
	\hat f(p_2)&=\frac{(1+p_2MS_{BC}/MS_{e_B})^2}{\frac{1}{(b-1)(r-1)}+\frac{(p_2MS_{BC}/MS_{e_B})^2}{(b-1)(c-1)}};
\end{align*}
finally, exchanging the order of $MS_1$ and $MS_2$, we obtain:
\begin{align*}
	p_2^*=\frac{(b-1)(r-1)}{(b-1)(r-1)-2}\left(\frac{2[(b-1)(r+c-2)-2]}{(b-1)(c-1)[(b-1)(r-1)-4]}+1\right),
\end{align*}
\begin{align*}
	\hat f(p_2^*)=\frac{(1+p_2^*MS_{e_B}/MS_{BC})^2}{\frac{1}{(b-1)(c-1)}+\frac{(p_2^*MS_{e_B}/MS_{BC})^2}{(b-1)(r-1)}}.
\end{align*}

\section{Application}

In \cite{Zimm}, a real example was considered when all the effects are fixed. The data in Tables \ref{tab6} and \ref{tab7} show the weight of 100 beans obtained by Luis Fernando Stone and Regis Vilela Bagatini on an experiment in 1998. It is a completely randomized design with two blocks on which each horizontal strip corresponds to the water layer irrigated, the vertical strips are soil tillage systems and the subplots are Nitrogen doses. The experiment was done at the Capivara farm in Embrapa Rice and Bean. 

\begin{table}[h]
\centering\small\caption{Block 1}\label{tab6}
\begin{tabular}{|c|c|c|c|}\hline
Water   &    Soil 1    & Soil 2      & Soil 3 \\ \hline
      & Nit 1 \vline Nit 2 \vline Nit 3 & Nit 1 \vline Nit 2 \vline Nit 3 & Nit 1 \vline Nit 2 \vline Nit 3\\
Water 1    & 26.33 \vline 27.85 \vline 27.13 & 25.10 \vline 27.67 \vline 24.93 & 25.00 \vline 28.03 \vline 29.65\\
Water 2    & 24.04 \vline 25.22 \vline 28.32 & 25.19 \vline 27.77 \vline 27.28 & 25.89 \vline 24.27 \vline 25.83\\
Water 3    & 25.85 \vline 25.70 \vline 26.97 & 25.63 \vline 27.11 \vline 25.62 & 26.16 \vline 24.86 \vline 25.51\\
Water 4    & 23.20 \vline 20.32 \vline 23.94 & 29.28 \vline 26.03 \vline 28.60 & 26.23 \vline 25.49 \vline 24.65\\ \hline
\end{tabular}
\end{table}
\begin{table}[htb]
\centering\small\caption{Block 2}\label{tab7}
\begin{tabular}{|c|c|c|c|}\hline
Water   &    Soil 1    & Soil 2      & Soil 3 \\ \hline
      & Nit 1 \vline Nit 2 \vline Nit 3 & Nit 1 \vline Nit 2 \vline Nit 3 & Nit 1 \vline Nit 2 \vline Nit 3 \\
Water 1    & 25.87 \vline 28.64 \vline 29.31 & 27.80 \vline 27.25 \vline 25.56 & 28.53 \vline 26.38 \vline 32.45\\
Water 2    & 27.16 \vline 26.49 \vline 25.99 & 24.63 \vline 26.91 \vline 28.47 & 26.68 \vline 27.64 \vline 24.80\\
Water 3    & 27.11 \vline 24.44 \vline 28.06 & 25.77 \vline 27.46 \vline 26.20 & 26.83 \vline 27.55 \vline 27.19\\
Water 4    & 23.00 \vline 23.43 \vline 23.42 & 28.71 \vline 26.45 \vline 26.25 & 26.64 \vline 26.82 \vline 26.88\\ \hline
\end{tabular}
\end{table}

The water layers (the vertical strips, $A$) are averaged irrigation levels as follows: 366.1 mm for the first horizontal strip, 335.1 mm for the second one, 315.7 mm for the third one, and 293.7 mm for the last one. There are three ways to prepare the soil (the vertical strips $B$): heavy harrowing for the first vertical strip, moldboard plowing for the second one, and notillage on the last one. The Nitrogen subdoses ($C$) inside the subplots are, respectively for each subplot, 0, 20 and 40 kg ha$^{-1}$.

The $MS$ and the $df$ needed to construct the $F$ tests are shown on Table \ref{tab8}. These results show that there are significant effects on the water layer, its interaction with the soil, and the interaction of the three factors.

\begin{table}[htb]
\centering\small\caption{Example}\label{tab8}
\begin{tabular}{|l|c|c|c|}\hline
Source   &  df &    $MS$    & 	$F$    \\ \hline
$R$      & $1$ & $9.4758$   & 				 \\
$A$      & $3$ & $10.9903$  & $26.04$  \\
$e_A$    & $3$ & $0.4220$   & 				 \\ \hline
$B$      & $2$ & $7.3937$   & $2.91$ 	 \\ 
$e_B$    & $2$ & $2.5387$   & 				 \\ \hline
$AB$     & $6$ & $11.2718$  & $35.89$  \\
$e_{AB}$ & $6$ & $0.3141$   & 				 \\ \hline
$C$      & $2$ & $3.1476$   & $2.11$ 	 \\ 
$AC$     & $6$ & $2.3759$   & $1.59$   \\ 
$BC$     & $4$ & $1.8678$   & $1.25$   \\ 
$ABC$    & $12$ & $3.2911$  & $2.21$   \\
$e_t$    & $24$ & $1.4921$	&					 \\ \hline
\end{tabular}
\end{table}

Now, in principle, it would be possible to analyze these data using other factorial experiments. Let us consider two of them: a three-way factorial experiment and split-split plot design. For these experiments, we present respectively Tables \ref{tab9} and \ref{tab10} below with the analysis of variance.

The first obvious thing is that the $B$ source (the soil) is not significative under the strip-split plot design, but is is significative under the two new models. It is important to note also the differences on the degrees of freedom for each model. In the strip-split plot design, for instance, the error of the effect $B$ is represented by the interaction of the blocks and the soil, with 2 degrees of freedom; in the case of the factorial design the experimental error has 35 degrees of freedom (assuming the fixed effects model); and in the split-split plot design the error has 8 degrees of freedom, since the interaction blocks-layer-soil also includes the blocks-soil interaction.

Now, these two cases are considered only for illustration purposes. However, it seems appropriate to remember that the design of the field experiment defines the mathematical model and, therefore, the ANOVA. The most common mistake in statistics analyses applied to agriculture is to disrespect this design. This simple exercise shows that it is possible to carry on several analyses for a given data set, but only one is correct ---the one following the field scheme. This field scheme is also determining the way to execute the randomization of treatments.

\begin{table}[h]
\centering\small\caption{Factorial design}\label{tab9}
\begin{tabular}{|l|c|c|c|}\hline
Source   &  df &    $MS$    & 	$F$    \\ \hline
$R$      & $1$ & $9.4758$   & 				 \\
$A$      & $3$ & $10.9903$  & $8.73$  \\
$B$      & $2$ & $7.3936$   & $5.88$ 	 \\ 
$AB$     & $6$ & $11.2718$  & $8.96$  \\
$C$      & $2$ & $3.1476$   & $2.50$ 	 \\ 
$AC$     & $6$ & $2.3759$   & $1.89$   \\ 
$BC$     & $4$ & $1.8678$   & $1.48$   \\ 
$ABC$    & $12$ & $3.2911$  & $2.62$   \\
$e_t$    & $35$ & $1.2582$	&					 \\ \hline
\end{tabular}
\end{table}

\begin{table}[h]
\centering\small\caption{Split-split plot design}\label{tab10}
\begin{tabular}{|l|c|c|c|}\hline
Source   &  df &    $MS$    & 	$F$    \\ \hline
$R$      & $1$ & $9.4758$   & 				 \\
$A$      & $3$ & $10.9903$  & $7.37$  \\
$e_A$    & $3$ & $0.4220$   & 				 \\ \hline
$B$      & $2$ & $7.3937$   & $4.96$ 	 \\ 
$AB$     & $6$ & $11.2718$  & $7.55$  \\
$e_{AB}$ & $8$ & $0.8702$   & 				 \\ \hline
$C$      & $2$ & $3.1476$   & $2.11$ 	 \\ 
$AC$     & $6$ & $2.3759$   & $1.59$   \\ 
$BC$     & $4$ & $1.8678$   & $1.25$   \\ 
$ABC$    & $12$ & $3.2911$  & $2.21$   \\
$e_t$    & $24$ & $1.4921$	&					 \\ \hline
\end{tabular}
\end{table}

\renewcommand{\refname}{References}

\end{document}